\documentclass[12pt,preprint]{aastex}

\newcommand{\msun}{M$_\odot$ }
\newcommand{\msyr}{M$_\odot$.yr$^{-1}$ } 
\newcommand{\al}{\mbox{$^{26}$Al}}

\newcommand{\clg}{\mbox{$^{34g}$Cl}}
\newcommand{\clm}{\mbox{$^{34m}$Cl}}

\def\power#1{\mbox{$\times10^{#1}\ $}}
\newcommand{\gap}{\mathrel{ \rlap{\raise.5ex\hbox{$>$}}
                      {\lower.5ex\hbox{$\sim$}}  } }
\newcommand{\lap}{\mathrel{ \rlap{\raise.5ex\hbox{$<$}}
		      {\lower.5ex\hbox{$\sim$}}  } }
\newcommand{\zaa}{Astron. Astrophys.}

\newcommand{\znp}{Nucl.~Phys.}

\newcommand{\zpr}{Phys.~Rev.}

\begin{document}

\title{\bf Synthesis of intermediate-mass elements in 
       classical novae: from Si to Ca}

\author{Jordi Jos\'e}
\affil{Departament de F\'{\i}sica i Enginyeria Nuclear (UPC), Av.
V\'{\i}ctor Balaguer, s/n, E-08800 Vilanova i la Geltr\'u (Barcelona),
SPAIN \\
and\\
Institut d'Estudis Espacials de Catalunya,
Edifici Nexus-201, C/ Gran Capit\`a 2-4, E-08034 Barcelona, SPAIN}
\author{Alain Coc}
\affil{Centre de Spectrom\'etrie Nucl\'eaire et de Spectrom\'etrie de
Masse, IN2P3-CNRS, Universit\'e Paris Sud, B\^at.104, F-91405 Orsay Campus, 
FRANCE} 
\and
\author{Margarita Hernanz}
\affil{Instituto de Ciencias del Espacio (CSIC), and\\ 
Institut d'Estudis Espacials de Catalunya,
Edifici Nexus-201, C/ Gran Capit\`a 2-4, E-08034 Barcelona, SPAIN}
	      
\received{}
\accepted{}

\slugcomment{\underline{Submitted to}: \apj~~~~\underline{Version}:
\today}

\begin{abstract}

 Thermonuclear runaways driven by accretion into degenerate
 white dwarf cores are the source that power classical nova outbursts.
 In this paper, we identify the dominant nuclear paths involved
 in the synthesis of intermediate-mass elements, from Si to Ca, during
 such violent events. 
 New evolutionary sequences of 1.35 \msun ONe novae have been computed,
 using updated nuclear reaction rates.
 The main nuclear activity in this region is powered
 by the leakage from the NeNa--MgAl region, where the activity is confined
 during the early stages of the explosion. We discuss the
 critical role played by $^{30}$P(p,$\gamma$) in the synthesis of 
 nuclear species beyond sulfur and point out the large uncertainty
 that affects its rate, which has dramatic consequences for studies
 of nova nucleosynthesis in the Si--Ca mass region.
\end{abstract}


\section{Introduction} 

 Classical novae inject nuclear processed
 material into the interstellar medium, and therefore may play a role in
 the Galactic nucleosynthesis. Isotopes likely attributed to 
 nova outbursts include $^{13}$C, $^{15}$N and $^{17}$O 
 (see Jos\'e \& Hernanz 1998, and references therein). 
 Current hydrodynamic models of classical novae show that explosions on
 carbon-oxygen white dwarfs (hereafter, CO novae) are
 dominated by the synthesis of CNO-group nuclei, largely overproduced
 in the ejecta with respect to solar abundances (Starrfield et al. 1997; 
 Kovetz \& Prialnik 1997; Jos\'e \& Hernanz 1998).
  On the other hand,
 explosions on oxygen-neon white dwarfs (hereafter, ONe novae), 
 more massive than CO ones,  
 provide large enrichments not only in CNO nuclei, but also in neon,
 sodium, aluminum, or other intermediate-mass elements
 (Starrfield et al. 1998; Jos\'e \& Hernanz 1998; Jos\'e et al. 1999).
 Hence, the composition of the ejecta is
 a key signature of the nature of the underlying white dwarf (CO or
 ONe) and provides also
 valuable information on the explosion (i.e., peak temperature,
 characteristic timescale...). In particular, the presence of isotopes of the 
 silicon to calcium mass region (hereafter, Si-Ca), 
 overproduced with respect to solar values,
 in the ejecta of a classical nova, may reveal the presence of an underlying
 massive white dwarf (the ones reaching the higher peak temperatures during
 the explosion) since proton captures on 
 nuclei above silicon require temperatures in excess of $3 \times 10^8$ K to 
 overcome their large Coulomb barriers. 
Indeed, abundance determinations from some 
 observed novae, such as V1370 Aql 1982 (Snijders et al. 1987),
 QU Vul 1984 (Andre\"a et al. 1994), 
 V2214 Oph 1988 (Andre\"a et al. 1994), or
  V838 Her 1991 (Vanlandingham et al. 1996, 1997), 
 show the presence of Si-Ca nuclei in the ejecta.

 Moreover, the nova contribution to the Galactic content of other species
 such as $^{7}$Li 
 (Starrfield et al. 1978, Hernanz et al. 1996) or $^{26}$Al 
 (Politano et al. 1995, Jos\'e et al. 1997, Starrfield et al. 1998, 
 Jos\'e et al. 1999) may also be significant, although its real extent is 
 still a matter of debate (see Romano et al. 1999, for details on
 the contribution of novae to match the $^{7}$Li content in realistic
 calculations of Galactic Chemical evolution). 
 The synthesis of such medium- or long-lived species has, in turn, implications 
 for the high-energy emission during nova outbursts: 
 several gamma-ray signatures have been
 predicted for nova outbursts (Clayton \& Hoyle 1974; Leising \& Clayton
 1987; G\'omez-Gomar et al. 1998,
 Hernanz et al. 1999), including a 478 keV line from 
 $^7$Li (CO novae), a 1275 keV line from $^{22}$Na (ONe novae), and an
 intense 511 keV line and the corresponding continuum at energies 
 below (for both CO and ONe nova types). In the Si-Ca mass region,
 another gamma-ray signal, coming from the $\beta^+$-decay of the medium
 lived $^{34m}$Cl (isomeric state), has also been proposed 
 for 511 keV (Leising \& Clayton 1987) and 2.128~MeV (Coc et al. 2000a) 
 line emission.
 However, no nucleosynthesis calculation with realistic nuclear reaction
 rates has, up to now, been attempted to evaluate the chances
 for the potential detection of this particular gamma-ray signal from novae. 

 Furthermore, the recent discovery of a number of presolar grains 
 likely formed in the ejecta of classical novae (Amari et al., 2001)
 also points out the importance of a good determination of the abundance
 pattern, in particular in the Si-Ca mass region: together with a number 
 of constraints
 from specific isotopic ratios (i.e., low $^{12}$C/$^{13}$C and 
 $^{14}$N/$^{15}$N values, high $^{26}$Al/$^{27}$Al), the basic feature 
 that points towards a nova origin relies on the large excesses 
 in $^{30}$Si/$^{28}$Si and close-to-solar $^{29}$Si/$^{28}$Si measured
 in such grains. The determination of other isotopic ratios in this
 mass range, such as 
 $^{33}$S/$^{32}$S, may also help to identify future nova grain candidates.  

These aspects point out the interest of a more accurate determination of
the relevant synthesis mechanisms in the Si-Ca mass region. Such an analysis
 has been only scarcely attempted (see recent
 parametrized nucleosynthesis calculations by Lefevbre et al. 1997,
 Wanajo et al. 1999, and Iliadis et al. 1999, as well as hydrodynamic
 calculations by Starrfield et al. 1998),  
  partially because of the large number of isotopes of interest in 
 this mass region, around 40 species, and the hundred nuclear reactions 
 involved. Accordingly, it is the purpose of this paper to 
 determine the dominant nuclear paths involved
 in the synthesis of intermediate-mass elements, from Si to Ca, during
 classical nova outbursts, identifying the key reactions and 
  investigating their associated
 nuclear uncertainties, that may have a dramatic influence in the overall
 process. In Section 2, we outline the basic nuclear paths at different stages
 of the nova outburst for a particular model. Details on nuclear physics
 aspects are given in Section 3. 
 Constraints on the production of Si-Ca group nuclei, together with
 the most relevant conclusions of this paper are summarized in Section 4.

\section{Main nuclear paths in the synthesis of Si-Ca nuclei 
           in classical novae}

 The high temperatures required to initiate a noticeable nuclear activity
 in the Si-Ca group nuclei restrict
 the domain of interest to explosions in very massive white dwarfs (i.e.,
 close to the Chandrasekhar mass). In this section, we will focus on the main
 nuclear paths involved in the synthesis of several isotopes within this mass
 range, through a detailed analysis of a 1.35 \msun ONe white dwarf, 
 which accretes solar-like matter at a rate of $\dot M = 2 \times 10^{-10}$
 \msyr, assuming a 50\% degree of mixing with the outermost shells of the
 white dwarf core. Initial abundances of the Si-Ca group nuclei adopted in
 this Model are listed in Table 1 (abundances up to Si are given in Jos\'e et
 al. 1999, Table 1). In fact, the specific prescription adopted for the
 initial abundance pattern is crucial for nucleosynthesis  calculations.
 Whereas a deep analysis of this effect is out of the scope of this paper,
 and will be presented elsewhere, we just point out that it introduces
 some uncertainty in all current nucleosynthesis calculations during nova
 outbursts.

The main source for the production of the Si--Ca isotopes (see Fig. 1) comes
from the processing of the initial Ne--Al isotopes from the white dwarf 
leaking out of the NeNa and MgAl mass regions, mainly through $^{28}$Si. 
We do not discuss here the nuclear processing within these regions but
refer the reader to a previous paper (Jos\'e et al. 1999). 
 Hence we start our
discussion with the last reactions linking the MgAl to SiP mass regions.
The new evolutionary sequences of nova outbursts have been
 calculated by means of the code SHIVA (see Jos\'e \& Hernanz 1998 for
 details), a one-dimensional implicit hydrodynamical code that follows
 the course of the explosion from the onset of accretion up to the ejection
 stages. Calculations have been performed with a nuclear
 reaction network updated for the region of interest (see
 section 3).  Snapshots of the evolution of the most relevant elements
 are shown in Figures 2 to 6.

 At the early stages of the explosion, nuclear activity above CNO cycle 
 is mostly confined in the NeNa and MgAl region with only
marginal activity in the Si--Ca region. 
 Soon after the thermonuclear runaway takes place, when the temperature at 
 the burning shell reaches 
 $T_{bs} = 8 \times 10^7$ K (Figs. 2 to 6, first panel),
 the main nuclear path affecting Si to Ca evolution is dominated by a
 series of reactions, beginning with proton captures onto $^{26g}$Al 
 (ground state), $^{26g}$Al(p,$\gamma$)$^{27}$Si, 
 followed by $^{27}$Si($\beta^+$)$^{27}$Al(p,$\gamma$)$^{28}$Si.
 As a result, $^{27}$Si is slightly enhanced at the burning shell, 
  since its $\beta^+$-decay channel is slower than $^{26g}$Al(p,$\gamma$).
 Marginal nuclear activity at the burning shell involves other proton-capture
 reactions on $^{25}$Al and $^{26m}$Al (isomeric state), as well as
 onto $^{27}$Si, which account for a moderate increase in $^{26}$Si 
 and $^{28}$P. 
 No relevant nuclear activity beyond phosphorus is found.

 Evolution at $T_{bs} = 10^8$ K  (Figs. 2 to 6, second panel) is 
 still dominated by the chain 
 $^{26g}$Al(p, $\gamma$)$^{27}$Si($\beta^+$)$^{27}$Al(p,$\gamma$)$^{28}$Si, 
 plus the contribution of 
 $^{25}$Al(p,$\gamma$)$^{26}$Si($\beta^+$)$^{26m}$Al(p,$\gamma$)$^{27}$Si, 
 and $^{27}$Si(p,$\gamma$)$^{28}$P($\beta^+$)$^{28}$Si. Some marginal
 activity involves also $^{31}$P(p,$\gamma$)$^{32}$S, and 
 $^{34}$S(p,$\gamma$)$^{35}$Cl.
 The most remarkable feature at this stage is an important increase in both 
 $^{26,27}$Si. 
 The outermost shells are dominated by some marginal decays (the lower 
 temperatures prevent proton captures to proceed efficiently), 
 such as $^{27}$Si($\beta^+$)$^{27}$Al, 
 $^{26}$Si($\beta^+$)$^{26m}$Al, and  $^{30}$P($\beta^+$)$^{30}$Si, followed
 by $^{29}$P($\beta^+$)$^{29}$Si, and $^{34m}$Cl($\beta^+$)$^{34}$S.

 Important changes take place as soon as $T_{bs}$ reaches $2 \times 10^8$ K 
 (Figs. 2 to 6, third panel).  The nuclear activity is now driven by 
 $^{26g}$Al(p,$\gamma$)$^{27}$Si(p,$\gamma$)$^{28}$P (faster than 
 $^{27}$Si($\beta^+$)$^{27}$Al), and by $^{27}$Al(p,$\gamma$)$^{28}$Si, 
 followed by 
 $^{25}$Al(p,$\gamma$)$^{26}$Si($\beta^+$)$^{26m}$Al(p,
                                     $\gamma$)$^{27}$Si($\beta^+$)$^{27}$Al,
 $^{28}$P($\beta^+$)$^{28}$Si, 
 $^{26}$Si(p,$\gamma$)$^{27}$P($\beta^+$)$^{27}$Si, 
 $^{28}$P(p,$\gamma$)$^{29}$S, and $^{28,29}$Si(p,$\gamma$)$^{29,30}$P.
 The outermost shells are dominated by $^{27}$Si($\beta^+$)$^{27}$Al, 
 $^{26}$Si($\beta^+$)$^{26m}$Al, and  $^{28}$P($\beta^+$)$^{28}$Si, followed
 by $^{27,29,30}$P($\beta^+$)$^{27,29,30}$Si.
 One of the most relevant outcomes is a substantial increase in 
 both $^{26,27}$Si which, except for $^{28}$Si, become the most abundant 
 silicon isotopes in the envelope now. 
 In fact $^{27}$Si attains its maximum value at this stage (soon after, 
 as the temperature rises towards $T_{peak}$, it begins to be destroyed
 by proton captures).  Concerning  $^{28}$Si,
 it has significantly increased in the burning shell, 
 up to $2.5 \times 10^{-3}$, by mass. On
 the contrary, $^{29}$Si has been slightly destroyed at the burning shell by 
 (p,$\gamma$) reactions, whereas $^{30}$Si remains mostly unaffected.
 Major activity involves also several phosphorus isotopes: 
 whereas $^{31}$P remains close to its
 initial abundance, major synthesis of $^{28}$P 
 (reaching $4.4 \times 10^{-4}$ by mass, at the burning
 shell) and $^{27,29,30}$P ($\sim 10^{-6}$, by mass) is also reported. 
 Only one sulfur isotope, $^{29}$S, shows a noticeable variation: it is, 
 in fact,
 increased by (p,$\gamma$) reactions on $^{28}$P nuclei at the burning shell, 
 and shortly after, carried away to the outer
  envelope by means of convective transport. 
 No significant nuclear path involving Cl, Ar, K or Ca is found yet.

 When $T_{bs}$ reaches $3 \times 10^8$ K (Figs. 2 to 6, fourth panel),
 the major nuclear activity is dominated by
 $^{28}$Si(p,$\gamma$)$^{29}$P(p,$\gamma$)$^{30}$S,
 $^{27}$Si(p,$\gamma$)$^{28}$P($\beta^+$)$^{28}$Si,
 $^{25}$Al(p,$\gamma$)$^{26}$Si(p,$\gamma$)$^{27}$P($\beta^+$)$^{27}$Si,
 $^{26m}$Al(p,$\gamma$)$^{27}$Si plus $^{30}$P(p,$\gamma$)$^{31}$S.
 One of the most relevant outcomes related to the synthesis of S to Ca
 isotopes is the key role played by $^{30}$P in the evolution: in
 fact, $^{30}$P(p,$\gamma$)$^{31}$S controls the path towards heavier
 species, partially because the path through $^{29}$P
 (i.e., $^{29}$P(p,$\gamma$)$^{30}$S(p,$\gamma$)$^{31}$Cl) is inhibited
 by the efficient inverse photodisintegration reaction
 $^{31}$Cl($\gamma$, p)$^{30}$S, which is rapidly followed by a $\beta^+$-decay,
 leading to $^{30}$P. 
 Hence $^{30}$P is a mandatory passing point to $^{32}$S via 
$^{30}$P(p,$\gamma)^{31}$S(p,$\gamma)^{32}$Cl($\beta+)^{32}$S or through
$^{30}$P(p,$\gamma)^{31}$S($\beta+)^{31}$P(p,$\gamma)^{32}$S, while
$^{30}$P($\beta+)^{30}$Si(p,$\gamma)^{31}$P(p,$\gamma)^{32}$S is strongly
inhibited by the slow $^{30}$P $\beta^+$-decay.

Other important reactions at this stage include
 $^{24,26g,27}$Al(p,$\gamma$)$^{25,27,28}$Si, $^{26}$Si($\beta^+$)$^{26m}$Al, 
 $^{28}$P(p,$\gamma$)$^{29}$S($\beta^+$)$^{29}$P($\beta^+$)$^{29}$Si(p,
 $\gamma$)$^{30}$P
 and $^{30,31}$S($\beta^+$)$^{30,31}$P. A minor role is also played by
 $^{25}$Si($\beta^+$)$^{25}$Al, $^{30}$P($\beta^+$)$^{30}$Si, 
 $^{30}$S(p,$\gamma$)$^{31}$Cl and its reverse reaction 
 $^{31}$Cl($\gamma$,p)$^{30}$S, and 
 both (p,$\gamma$) and (p,$\alpha$) reactions on $^{31}$P.
 The outermost envelope, much cooler, is dominated by 
 $\beta^+$-decays from
 $^{26,27}$Si, $^{29,30}$P and marginally from $^{28}$P and $^{37}$K.
 The number of nuclei involved in the different dominant paths is, at this 
 stage, very large: the 
 most abundant silicon isotope, $^{28}$Si, has already reached 
 $3.3 \times 10^{-2}$ in the inner envelope.
 $^{26}$Si attains now its maximum value in the envelope. Regarding
 $^{27}$Si, it shows a peculiar profile, with a 'hump' at 
intermediate shells. This reflects the fact that, whereas it is destroyed 
 at the burning shell (where (p,$\gamma$) reactions dominate synthesis from 
 $^{27}$P($\beta^+$)) and at the outer part of the envelope 
 (through $\beta^+$-decays), it is enhanced by 
 $^{26g}$Al(p,$\gamma$) at such intermediate shells. A similar
 trend is found for $^{29}$Si: destroyed by 
 (p,$\gamma$) reactions in the inner envelope, and synthesized by
  $^{29}$P($\beta^+$) at intermediate shells. 
 $^{30}$Si increases only by a small amount.

 Concerning phosphorus, all isotopes are enhanced at this stage: 
 $^{27,28,30}$P reach $\sim 10^{-3}$, by mass, whereas $^{29}$P, the most 
 abundant phosphorus isotope so far, achieves $\sim 10^{-2}$. 
 $^{31}$P reaches $\sim 10^{-4}$ at the burning shell.
 Except for $^{37}$Cl (slightly destroyed at the burning shell), all other 
 chlorine isotopes are enhanced. 
 The most abundant one, $^{35}$Cl, reaches a mass-fraction 
 of $5.6 \times 10^{-6}$. The profile of $^{31}$Cl shows also a hump at 
 intermediate shells, which distinguishes the region where destruction by
 $^{31}$Cl(p,$\gamma$) dominates synthesis by $^{30}$S(p,$\gamma$)$^{31}$Cl 
 (innermost envelope) from that where synthesis by 
 $^{30}$S(p,$\gamma$)$^{31}$Cl dominates destruction by both 
 proton captures and $\beta^+$-decays (the evolution of $^{31}$Cl in the 
 outermost envelope is dominated by $^{31}$Cl($\beta^+$)). 
 
 Evolution of sulfur isotopes mainly affects $^{29-31}$S, which show
 a dramatic increase powered by (p,$\gamma$) reactions on $^{28-30}$P.
 In fact, $^{30}$S becomes the most abundant sulfur isotope at this stage.
 Moderate enhancement of $^{32}$S and destruction of $^{33,34,36}$S is
 also reported. 
 Some marginal activity involves also Ar and K, mainly affecting 
 $^{34,35,37}$Ar, enhanced at the burning shell, and $^{38}$Ar, which
 moderately decreases. 

 The course of the thermonuclear runaway drives the burning shell up to a peak 
 temperature of $T_{max} = 3.26 \times 10^8$
 K (Figs. 2 to 6, panel fifth). The main nuclear activity is now driven 
 by $^{25}$Al(p,$\gamma$)$^{26}$Si(p,$\gamma$)$^{27}$P and the chain
 $^{27}$Si(p,$\gamma$)$^{28}$P($\beta^+$)$^{28}$Si(p,$\gamma$)$^{29}$P(p,
 $\gamma$)$^{30}$S($\beta^+$)$^{30}$P(p,$\gamma$)$^{31}$S($\beta^+$)$^{31}$P(p,
 $\gamma$)$^{32}$S.
  Significant activity also involves 
 $^{24}$Al(p,$\gamma$)$^{25}$Si($\beta^+$)$^{25}$Al,
 $^{26}$Si($\beta^+$)$^{26m}$Al(p,$\gamma$)$^{27}$Si, 
 $^{26g}$Al(p,$\gamma$)$^{27}$Si, 
 $^{27}$P($\beta^+$)$^{27}$Si, 
 $^{28}$P(p,$\gamma$)$^{29}$S($\beta^+$)$^{29}$P($\beta^+$)$^{29}$Si(p,
 $\gamma$)$^{30}$P, \newline $^{31}$P(p,$\alpha$)$^{28}$Si and
 $^{30}$S(p,$\gamma$)$^{31}$Cl and its reverse photodisintegration reaction.
 At this stage, protons are energetic enough to overcome the Coulomb potential 
 barrier of many nuclei, in particular 
 $^{28}$Si which, for the first time, begins to decrease at the burning 
 shell. Reduction of  
 $^{26,27}$Si and a noticeable enhancement of $^{29,30}$Si is also found.
 Concerning phosphorus, $^{29,30}$P but mainly $^{31}$P increase at this
 stage. $\beta^+$-decays are responsible for a significant reduction of
 both $^{27,28}$P. The most relevant feature regarding the evolution of
 sulfur isotopes is an important increase of $^{30-33}$S. In fact, at this 
  stage, $^{31}$S becomes the most abundant sulfur isotope in the envelope.
 $^{33}$Cl is favored also by the rise in temperature, and becomes now the
 most abundant chlorine isotope in the vicinity of the burning shell. 
 $^{32,34g,34m}$Cl are also enhanced during this stage.
 Nuclear activity affects now the edge of the network, with some enhancement
 of heavier species, such as $^{34,35,37}$Ar, $^{37,38}$K and, to some extent,
 $^{39}$Ca. However, the moderately low nuclear reaction fluxes in this region
 justifies our choice of $^{40}$Ca as the last isotope of the network.  
 As earlier in the evolution of the nova outburst, the outer envelope is 
 again dominated by $\beta^+$-decays such as 
 $^{26,27}$Si($\beta^+$)$^{26m,27}$Al and $^{29,30}$P($\beta^+$)$^{29,30}$Si.

 Due to the sudden release of energy from the short-lived $\beta^+$-unstable
 nuclei $^{13}$N, $^{14,15}$O and $^{17}$F, the envelope begins to expand.
 A hundred seconds after $T_{peak}$, the envelope reaches a size of $10^9$ cm
 (Figs. 2 to 6, sixth panel). At this stage, the burning shell is dominated 
 by $^{26g}$Al(p,$\gamma$)$^{27}$Si(p, 
         $\gamma$)$^{28}$P($\beta^+$)$^{28}$Si(p,$\gamma$)$^{29}$P, 
         $^{29}$Si(p,$\gamma$)$^{30}$P(p,$\gamma$)$^{31}$S, and
         $^{31}$P(p,$\gamma$)$^{32}$S,
 plus a significant contribution from 
 $^{25}$Al(p,$\gamma$)$^{26}$Si($\beta^+$)$^{26m}$Al(p,$\gamma$)$^{27}$Si,
 $^{27}$Si($\beta^+$)$^{27}$Al(p,$\gamma$)$^{28}$Si, 
 $^{29}$P($\beta^+$)$^{29}$Si, 
 $^{29}$P(p,$\gamma$)$^{30}$S($\beta^+$)$^{30}$P($\beta^+$)$^{30}$Si(p,
     $\gamma$)$^{31}$P(p,$\alpha$)$^{28}$Si, 
 $^{31}$S($\beta^+$)$^{31}$P, and $^{32,34}$S(p,$\gamma$)$^{33,35}$Cl.
 The expansion of the envelope, which is accompanied by a drop in temperature,
 provides a quite different chemical pattern above the burning shell.
 In particular, the evolution of the Si-Ca nuclei is fully dominated 
 by $\beta^+$-decays (the most important ones
 being 
 $^{28-30}$P($\beta^+$)$^{28-30}$Si, $^{27}$Si($\beta^+$)$^{27}$Al,
 and $^{31}$S($\beta^+$)$^{31}$P, at the inner shells, whereas the
 outer envelope is dominated by 
 $^{29,30}$P($\beta^+$)$^{29,30}$Si, $^{34m}$Cl($\beta^+$)$^{34}$S,
 and $^{38}$K($\beta^+$)$^{38}$Ar). 
 At this stage of the evolution, the envelope shows dramatic changes
 in composition. In particular, isotopes such as $^{28,29,30}$Si, 
 $^{30,31}$P, $^{32-34}$S,
 $^{35}$Cl, $^{36,37}$Ar or $^{38}$K show a significant enhancement
 whereas 
 $^{26}$Si, $^{29-31}$S, $^{27-29}$P, $^{31-34}$Cl, 
  or $^{34}$Ar are efficiently destroyed.

 During the final stages of the outburst (Figs. 2 to 6, seventh panel),
 convection recedes from the surface and the burning shell becomes
 detached from the major part of the envelope, which is ultimately ejected.
 When the envelope reaches a size of $10^{10}$ cm, the main nuclear
 path in the Si-Ca region is basically carried by 
 $^{28-30}$P($\beta^+$)$^{28-30}$Si, $^{26,27}$Si($\beta^+$)$^{26m,27}$Al,
 $^{30,31}$S($\beta^+$)$^{30,31}$P, and $^{33}$Cl($\beta^+$)$^{33}$S,
 with $^{30}$P($\beta^+$)$^{30}$Si being the dominant one within most of 
 the envelope (again, contribution from 
 $^{34m}$Cl($\beta^+$)$^{34}$S and $^{38}$K($\beta^+$)$^{38}$Ar are
 important in the outer envelope).

 Soon after, when the envelope reaches a size of $10^{12}$ cm
 (Figs. 2 to 6, eighth panel), most of the 
 $\beta^+$-unstable nuclei have already decayed. Only the medium-lived ones 
 ($\tau \sim$ minutes), contribute yet to the energy production:
$^{30}$P($\beta^+$)$^{30}$Si,
 $^{34m}$Cl($\beta^+$)$^{34}$S, and $^{38}$K($\beta^+$)$^{38}$Ar. 
 The mean composition of the
 ejecta in the Si-Ca mass region consists, for this particular model,
  mainly of $^{28-30}$Si, $^{32}$S, and $^{31}$P. 
 The most overproduced species with respect to solar are $^{31}$P and
 $^{30}$Si, with overproduction factors (i.e., ratios between mean
 mass fractions in the ejecta over solar values) 1100 and 600,
 respectively. $^{32,33}$S are overproduced by a factor $\sim$ 100. 

 In summary, nuclear activity in the Si--Ca mass region is powered
 by leakage from the NeNa--MgAl region, where the activity is confined
 during the early stages of the ourburst. The main nuclear reaction
 that drives nuclear activity towards heavier species beyond sulfur  
 is $^{30}$P(p,$\gamma$), through two different paths: either
 $^{30}$P(p,$\gamma)^{31}$S(p,$\gamma)^{32}$Cl($\beta+)^{32}$S or
$^{30}$P(p,$\gamma)^{31}$S($\beta+)^{31}$P(p,$\gamma)^{32}$S. Then,
 the nuclear flow is dramatically reduced by the slow $^{32}$S(p,$\gamma$)
 limiting the production of heavier isotopes in the S-Ca mass range.

\section{Nuclear reaction rates in the Si-Ca mass region}

\subsection{Proton captures on silicon isotopes}

The rates of the $^{26,27}$Si(p,$\gamma$) reactions have been discussed in
Jos\'e et al. (1999). These reactions do not play a dominant role in
the path towards heavier species as
photodisintegration of $^{27}$P is effective (low Q-value) and because the
$^{27,28}$P lifetimes are small ($<$ 1~s).
The $^{28}$Si(p,$\gamma)^{29}$P reaction is the only significant
leak from the Ne--Al region towards heavier nuclei (Figure 1).
Hence it is natural to begin our analysis at this point, referring for the
discussion of rates in the Ne--Al chains to our previous paper (Jos\'e et al.
1999).
Except for the very recent work of Iliadis et al. (2001), published after this
work was completed, the two 
last compilations of thermonuclear rates end at silicon, either with
$^{30}$Si(p,$\gamma)^{31}$P (Caughlan \& Fowler 1988, hereafter CF88) or
$^{28}$Si(p,$\gamma)^{29}$P (Angulo et al. 1999, hereafter NACRE).
The $^{28}$Si(p,$\gamma)^{29}$P rate from NACRE differs significantly
from the CF88 one but only below 10$^8$~K, while the $^{29,30}$Si(p,$\gamma$) 
reaction rates should not deviate much from the CF88 ones at nova 
temperatures (Angulo, 2000). Accordingly, in this paper we have used the 
$^{28-30}$Si(p,$\gamma$) rates from CF88.

\subsection{Proton captures on phosphorus isotopes}

The $^{29}$P(p,$\gamma)^{30}$S rate was calculated by Wiescher \&
G\"orres (1988) from the properties of the T=1 isospin triplet.
The positions of the first levels above threshold in $^{30}$S were 
calculated by the Thomas--Ehrman shift method while the spectroscopic factors
were taken from the analog levels in $^{30}$Si. 
This rate is for the moment rather poorly known at nova temperatures 
as a consequence, in particular, of the uncertainty on the calculated 
position of the first resonance. 
However, this reaction rate is not crucial as there exists an alternative link
to heavier isotopes initiated by the $^{29}$P $\beta^+$-decay (Figure 1).
For the $^{30}$P(p,$\gamma)^{31}$S rate
(6.0371D+01, 8.6903D-01,-5.3250D+01, -4.5676D+00,
1.7458D+00,-1.5449D-01,-1.1437D+01, in REACLIB format),               
 we use the result of a statistical 
model calculation (Thielemann et al. 1987, hereafter SMOKER), because of
the lack of spectroscopic information on $^{31}$S.
Indeed, even though 17 levels are known up to 1 MeV above threshold,
very few spins are reported. Hence, it is not possible to calculate a realistic
reaction rate based neither on $^{31}$P analog levels nor shell model
calculations (Endt 1998). However,
the statistical model assumes a high level density which is {\it a priori}
favored by the high Q-value (6.133~MeV) but remains questionable at the moderate
temperatures found in novae.
Hence, with such a level density, it is more likely that the rate is 
governed by the properties of a few individual levels.

The inspection of Figure 1, shows that this reaction plays a key role as
$^{30}$P is a mandatory passing point to reach heavier nuclei and with its 
relatively long lifetime ($t_{1/2}$=2.5~mn) it stops further nucleosynthesis
unless proton captures are fast enough. 
Accordingly we investigate the effect of a possible
change of this rate by a test factor of 100 (see Section 4). 
The $^{31}$P(p,$\gamma)^{32}$S and $^{31}$P(p,$\alpha)^{28}$Si reaction rates
are deduced from experimental results (Iliadis et al. 1991; 1993; Ross et al. 
1995). 
While the first one
is well known, the second still suffers from uncertainties ($\lap$100) between
$T_9$ = 0.1 and 0.2 due to unknown resonance strengths. 
However, even the upper limit of the $^{31}$P(p,$\alpha)^{28}$Si rate remains
smaller than the $^{31}$P(p,$\gamma)^{32}$S one at nova temperature.
Hence this uncertainty should scarcely affect $^{31}$P destruction and 
nucleosynthesis beyond phosphorus.
 
\subsection{Proton captures on sulfur isotopes and beyond}

For $^{30}$S(p,$\gamma$)$^{31}$Cl, we use the rate calculated by Herndl et al. (1995)
 from shell model calculations. Another rate was provided by Van~Wormer 
et al. (1994) based on the properties of the analog levels in $^{31}$Cl. 
The two rates differ by a factor of $\lap$30 (Fig. 1 in Herndl et 
al. (1995)) due to the unknown location of the first excited
state in $^{31}$Cl. 
This should however have little effect since photodisintegration 
of $^{31}$Cl is fast (Q=0.286 MeV).
The $^{31}$S(p,$\gamma)^{32}$Cl reaction rate was first studied by Vouzoukas
et al. (1994). They measured the energy of the relevant $^{32}$Cl levels
via a $^{32}$S($^3$He,$t)^{32}$Cl transfer reaction. They used experimental 
spectroscopic factors from the $^{32}$P analog levels and radiative widths
from shell model calculations and their calculated ratios
($\Gamma_\gamma/\Gamma_T$)
are compatible with the Lefevbre et al. (1997) measurements.
A detailed investigation of the uncertainty affecting this rate has been performed
recently by Iliadis et al. (1999) leading to a nominal rate (adopted here) slightly lower 
than the
previous one (Vouzoukas et al. 1994), and a factor of 3 uncertainty in the 
domain of nova nucleosynthesis. One can thus consider this rate as 
sufficiently known.
The $^{32}$S(p,$\gamma)^{33}$Cl reaction rate was calculated by 
Iliadis et al. (1992) from  measured resonance strengths or extracted
spectroscopic factors (see also Iliadis et al. 1999, footnote in Table 7). The
 uncertainty on this rate was found to be very small (Thompson \& Iliadis 1999).
It is important to notice that the rate deduced from experimental data is much
smaller (see Iliadis et al. 1992) than the previously adopted reaction rate. 

For the reactions $^{33}$S(p,$\gamma$)$^{34g,34m}$Cl, we have taken into 
account the formation of both the $^{34}$Cl ground ($t_{1/2}$ = 1.53~s) and 
isomeric ($t_{1/2}$ = 32.0 min) states which, as in \al, 
have to be treated as separate isotopes (Coc et al. 2000a) 
at nova temperatures. 
We considered the resonances corresponding to the first five levels above
threshold (Endt 1990). Their branching ratios to the ground
or isomeric states were extracted from Endt (1990).
For the four unknown resonance strengths, we used upper limits with the usual
0.1 reduction factor. They were obtained from estimates of the proton widths
($\Gamma_p\ll\Gamma_\gamma$) based on calculation of single particle in a
Wood-Saxon potential whose depth is adjusted so as to reproduce the excitation
energy of the resonance.
The corresponding estimated $^{33}$S(p,$\gamma)^{34m}$Cl rate is given by:
\begin{eqnarray*}
N_{\mathrm A} \langle \sigma v \rangle=
& 1.5923\power{5}\,T_9^{-3/2}\times(0\;{\mathrm to}\;1)\times \\
                 & (2.4\power{-29}\exp\left(-0.3389\,/T_9\right)\times1.\\
                 & +1.5\power{-13}\exp\left(-2.003\,/T_9\right)\times0.89\\
                 & +2.4\power{-5}\exp\left(-2.836\,/T_9\right)\times0.64\\
                 & +1.\power{-3}\exp\left(-4.623\,/T_9\right)\times0.76)\\
   & +2.13\power{4}\,T_9^{-0.1493}\exp\left(-4.814\,/T_9\right) \times 0.5\\
\end{eqnarray*}
where the multiplying factors (1., 0.89, 0.64, 0.76 and 0.5) correspond to the
branching ratios to the isomeric state (Endt 1990). The
$^{33}$S(p,$\gamma)^{34g}$Cl is hence given by replacing these factors by
(0., 0.11, 0.36, 0.24 and 0.5). The last term is a multiresonant 
contribution from the higher energy resonances (Endt 1990) with an assumed 
average branching ratio of 0.5. 

Formation of \clm\ proceeds only through the
$^{33}$S(p,$\gamma)^{34m}$Cl reaction since, after internal transitions,
$^{34}$Ar $\beta^+$-decays lead only to \clg\ (a similar
configuration is found for $^{26}$Al). An important aspect of
 \clm\ nucleosynthesis is its destruction by
thermally induced radiative transitions to its short lived ground state. The
corresponding effective lifetime, that we use here, has been calculated
(Shell Model) by Coc et al. (2000a): it exhibits a rapid evolution (3 orders
of magnitude)
from its laboratory half-life value of 32.0~min down to $\approx$1~s.
This transition occurs 
around T$_9\approx$0.15-0.2, a typical domain of temperatures in
novae, and this strongly modifies the prospect for its
detectability through its gamma-ray signature. 
Other nuclei, such as $^{24}$Al, 
have also isomers in the mass region of nova
nucleosynthesis but they are at the fringe of nuclear flow
and their lifetimes are much less affected than $^{34}$Cl at nova
temperatures.

With some notable exceptions like $^{35}$Cl(p,$\gamma$) and 
$^{36}$Ar(p,$\gamma$) (see also the very recent compilation of 
Iliadis et al. 2001), experimental and theoretical (i.e., 
Shell Model, see Herndl et al. 1995) data 
are in general not sufficient to allow the explicit calculation of the rates at
nova temperatures for reactions beyond sulfur. Accordingly,  we use the
SMOKER rates instead (Thielemann et al. 1987) which are virtually identical to 
the more recent MOST rates from Goriely (1996), for these nuclei close to 
stability.

\section{Results and discussion}

 Table 2 shows the mean composition of the ejecta, for the Si-Ca mass
 region, corresponding to the 1.35 \msun ONe nova described in Section 2
 (hereafter, Model 135nom).
 The most abundant nuclei, for this specific mass range, are $^{32}$S,
 $^{28}$Si, $^{30}$Si, $^{31}$P and $^{29}$Si. For the purpose of comparison,
 a very similar model (i.e., Model ONe135B. See Jos\'e, Coc \& Hernanz 1999),
 computed with a previous nuclear reaction network, is also shown in Table 2.
 Both nuclear reaction networks are very similar up to silicon. Recent updates
 included in the new computations involve
 $^{18}$F(p,$\gamma$), $^{18}$F(p,$\alpha$) (Coc et al. 2000b, nominal rates),
 $^{17}$O(p,$\gamma$), $^{17}$O(p,$\alpha$) (NACRE, nominal rates),
 and also $^{31}$P(p,$\gamma$), $^{31}$P(p,$\alpha$),
 $^{31}$S(p,$\gamma$), $^{32}$S(p,$\gamma$),
 $^{33}$S(p,$\gamma$) (leading to both $^{34g,34m}$Cl),
and the corresponding $^{34g,34m}$Cl, temperature-dependent
$\beta^+$-decays (see Section 3 for specific references).
 As expected, no significant differences are found for Si or P yields
 when we switch to the updated network since, 
 as discussed in Coc et al. (2000b), proton captures onto $^{17}$O or $^{18}$F
 don't play a dominant role in the energetics of the explosion, and hence,
 a similar peak temperature and characteristic timescales for the evolution
 are found.
 Differences, however, are found for $^{32}$S and beyond.
 In fact, the new $^{32}$S yield increases by a factor of 2, because
 of the larger new $^{31}$P(p,$\gamma$) rate (below $10^8$ K and  also in the
 vicinity of $T_{peak}$) and the lower $^{32}$S(p,$\gamma)^{33}$Cl destruction 
rate.
 On the contrary, $^{33,34}$S, $^{35,37}$Cl and $^{36,38}$Ar
 are dramatically reduced by, at least, a factor of 10. This remarkable change
 results mainly from the update of the $^{32}$S(p,$\gamma$) rate, much lower 
 than the previous one in the domain of nova temperatures, 
 as pointed out by Iliadis et al. (1992). 
 Hence, whereas the arguments to invoke a likely nova origin for a number
 of presolar meteoritic grains (see Amari et al. 2001) remain unaffected by
 the results reported in this paper (i.e., silicon excesses), the differences 
 found for other
 nuclear species, in particular the $^{32}$S/$^{33}$S ratios, are relevant
 for the unambiguous identification of future nova grain candidates: 
 for instance, the previous $^{32}$S/$^{33}$S isotopic ratio 
 of 3.1, found for Model ONe135B, has now raised up to 97 for Model 135nom, 
 basically due to the update of the $^{31}$P(p,$\gamma$) and 
 $^{32}$S(p,$\gamma$) rates. 

 It is also worth noting that our numerical calculations reveal the presence of 
 the gamma-ray emitter $^{34m}$Cl in the ejected envelope. However,
 its content has been dramatically reduced by the update of the 
 $^{32}$S(p,$\gamma$)$^{33}$Cl and $^{33}$S(p,$\gamma$)$^{34m,34g}$Cl 
rates, together with the $^{34m}$Cl $\beta^+$-decay:
 whereas a mean mass fraction of $3.4 \times 10^{-3}$, 30 minutes after
 $T_{peak}$, was found for Model ONe135B, only $7 \times 10^{-7}$ by mass,
 remain, for the same time, in the present Model 135nom. Hence, the possibility of a
 potential detection of the gamma ray emission 
 corresponding to $^{34m}$Cl $\beta^+$-decay, 
 invoked as another signature of a classical novae
 (see Leising \& Clayton 1987), seems to be ruled out with the
 present calculations. 

 As pointed out, one of the most relevant outcomes of this study, 
 from the nuclear physics viewpoint,  is the crucial role played
 by $^{30}$P(p,$\gamma$) in the synthesis of heavier nuclear species. 
 Its purely theoretical rate is questionable since the model used for
its calculation reaches its limits for such a light nucleus at low temperature.
 Hence, we have estimated its impact
 on the yields by computing two new evolutionary sequences of nova models, 
 for which we have arbitrarily modified the $^{30}$P(p,$\gamma$) rate: 
 Model p30low, where the $^{30}$P(p,$\gamma$) has been reduced
 by a test factor of 100 throughout the calculation, and Model p30high, where 
 the nominal rate has been arbitrarily multiplied by 100. Results are summarized 
 also in Table 2. The increase by a factor of 100  
 affects only the $^{30}$Si yield: since the path towards $^{31}$S is favored 
 (which in turn, accounts for the moderate increase of $^{31}$P in the ejecta),
 much less $^{30}$Si (coming from $^{30}$P($\beta^+$)) is left in the envelope. 
 However, the impact on other
 nuclear species is very limited. On the contrary, when the 
 $^{30}$P(p,$\gamma$) rate 
 is reduced by a factor of 100, dramatic changes in the expected yields occur.
 Now, $^{30}$P($\beta^+$) competes favorably with proton captures, and an 
 important
 enhancement of $^{30}$Si is obtained. In fact, the reduction of the proton 
 capture rate halts significantly the synthesis of elements above Si (by a 
 factor of $\sim$ 10 with respect to the values found with the nominal rate). 
 We have compared our results with the latest available results from hydrodynamic
 simulations of ONe novae by Starrfield et al. (2001): they  
 state that the switch to an updated nuclear network (Iliadis et al. 2001) 
 is accompanied by an 
 important reduction of the abundances of nuclei above aluminum, in good
 agreement with the results found here.
 However, our new computations show that reduction of the yields affect
 only nuclei above $^{32}$S. In fact, a significant reduction of $^{32}$S 
 results only when we adopt the low $^{30}$P(p,$\gamma$) test rate. 
 Hence, this discrepancy might be likely attributed to 
 a different prescription for the adopted $^{30}$P(p,$\gamma$) rate. In
 particular, the rate  used by Starrfield et al. (2001) seems to be
  somewhat lower than the nominal rate adopted in 
 our paper, which stresses again the crucial dependence of the S--Ca yields on 
 this particular rate.

 Despite a detailed comparison with observations
 is out of the scope of this paper, because a larger number of nova
 models is required to properly explore the wide parameter space, 
  some conclusions can be outlined. 
 In fact, spectroscopic abundance determinations
 show evidence for some overproduction of nuclei in the Si--Ca mass region
 in a number of classical novae, including silicon
 (Nova Aql 1982, Snijders et al. 1987, Andre\"a et al. 1994; QU Vul 1984,
  Andre\"a et al. 1994), sulfur (Nova Aql 1982, Snijders et al. 1987, 
  Andre\"a et al. 1994), chlorine (Nova GQ Mus 1983, Morisset \& Pequignot 1996),
 argon and calcium (Nova GQ Mus 1983, Morisset \& Pequignot 1996;
  Nova V2214 Oph 1988, Nova V977 Sco 1989 and Nova V443 Sct 1989, Andre\"a et
  al. 1994). Indeed, models of explosions in 1.35 \msun ONe white dwarfs have an 
 end-point for nucleosynthesis below calcium, hence in agreement
with observations. This suggests that peak temperatures attained at
  the burning shell during nova outbursts cannot be much larger than  
  $3 \times 10^8$ K, otherwise an rp-process would result at such high 
  temperatures, driving the nuclear path towards heavier species, not
  observed so far. 
  Moreover, as stated in the discussion of the main nuclear paths, 
 determination of chemical abundances in the Si-Ca mass region  
 during nova outbursts is affected by uncertainties of nuclear physics
 origin. In particular, the uncertainty associated to  
 $^{30}$P(p,$\gamma$) makes it difficult to determine
 realistic abundances above silicon, and to clarify as well 
 whether the path towards
 heavier species is halted or, on
 the contrary, if significant production of Si-Ca nuclei is still expected.
 A better determination of this crucial rate would be of significant importance
 for the studies of nova nucleosynthesis that takes place in massive ONe white
 dwarfs.
   
\acknowledgments
The authors would like to thank an anonymous referee for his suggestions.
This research has been partially supported by the CICYT-P.N.I.E.
(ESP98-1348), by the DGICYT (PB97-0983-C03-02; PB97-0983-C03-03), 
 and by the MCYT (AI HF1999-0140).

\clearpage

\begin{figure}
\plotone{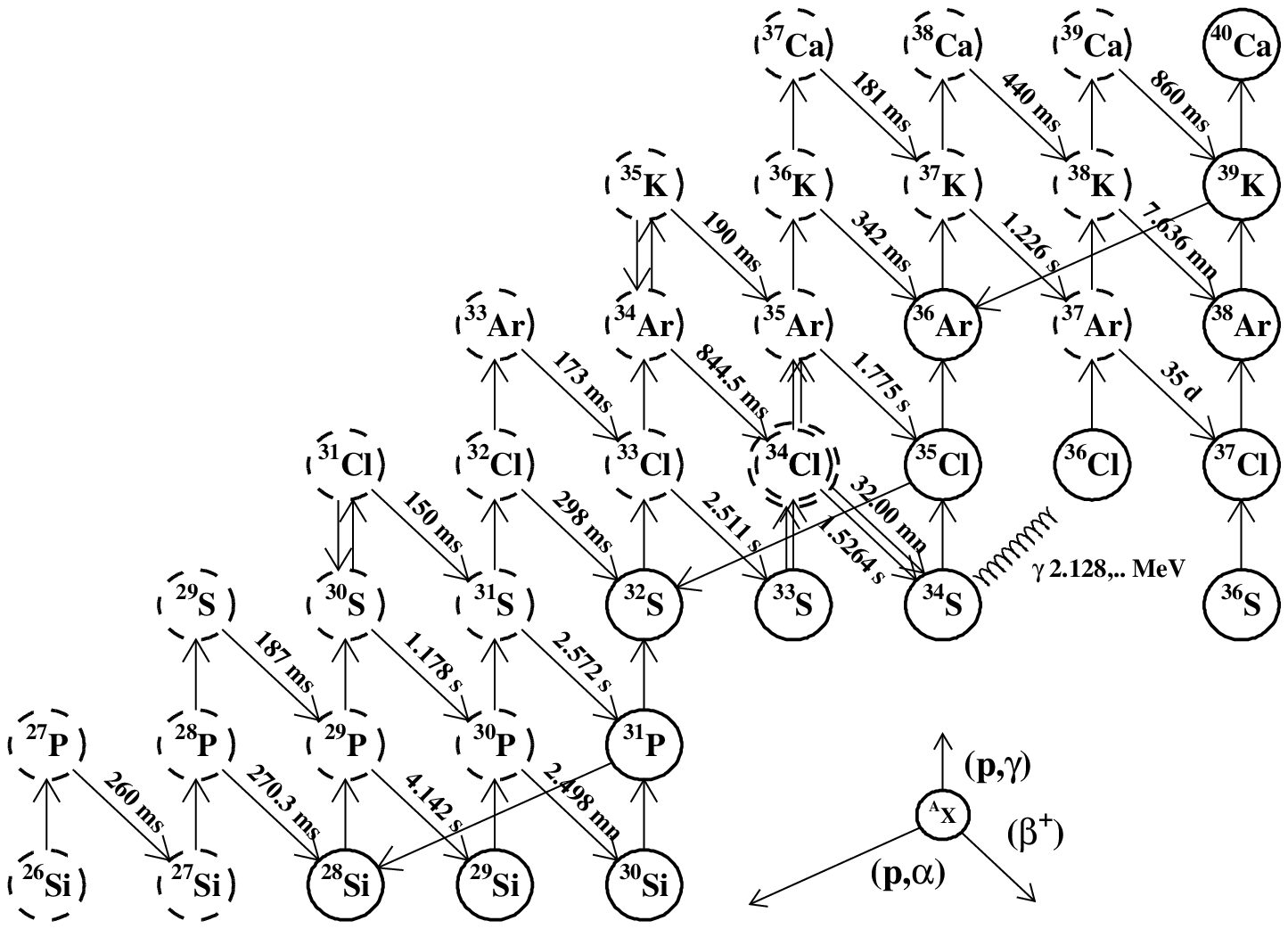}
\caption{Main nuclear paths in the Si-Ca mass region. \label{fig1}}
\end{figure}

\clearpage

\begin{figure}
\plotone{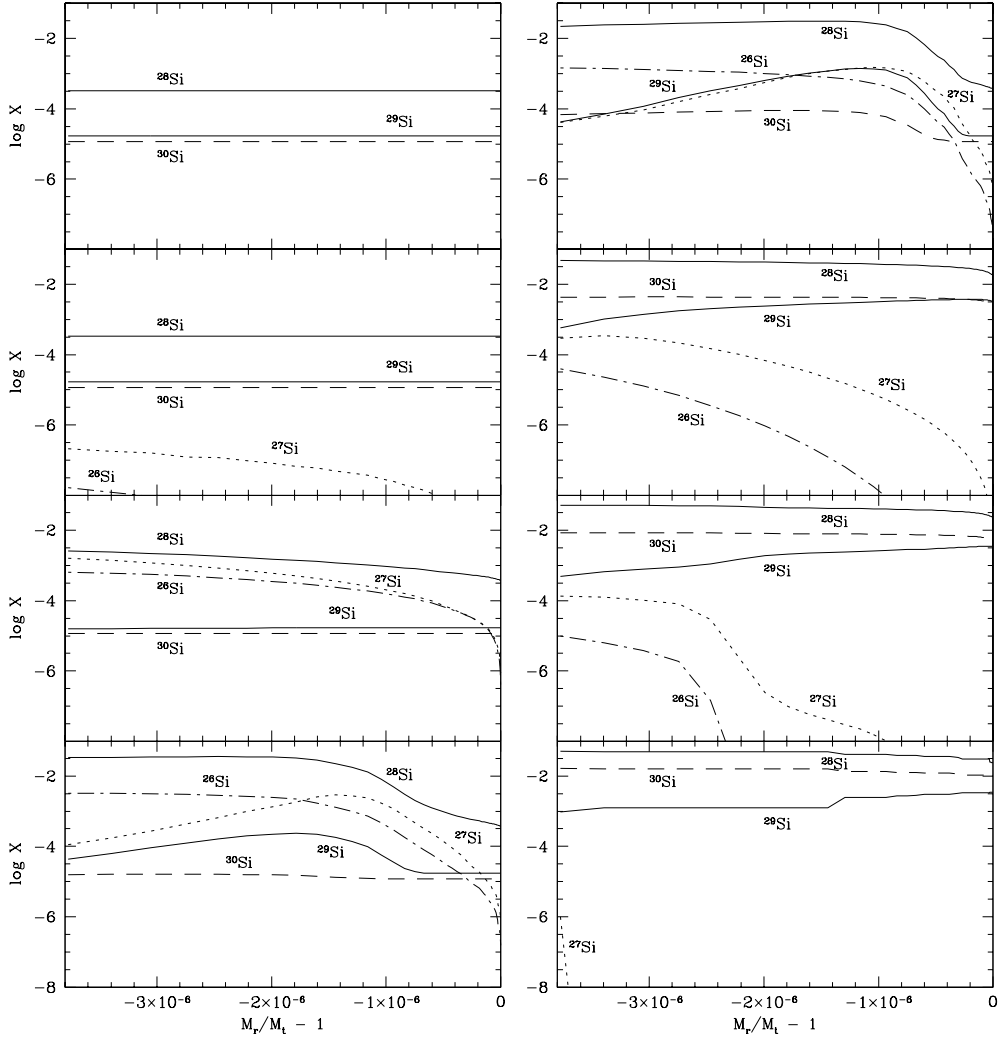}
\caption{Snapshots of the evolution of silicon isotopes (in mass
 fractions) along the accreted envelope, for a 1.35 \msun ONe nova accreting
 at a rate of $\dot M = 2 \times 10^{-10}$ \msyr. The mass coordinate
 represents the mass below the surface relative to the total mass. From top
 to bottom, panels correspond to the time for which the temperature at the
 burning shell reaches $8 \times 10^7$, $10^8$, $2 \times 10^8$, $3 \times
 10^8$, and $T_{peak} = 3.26 \times 10^8$ K, plus three panels corresponding
 to the last phases of the explosion, when the white dwarf envelope 
 has already expanded to a size of $R_{wd} \sim 10^9, 10^{10}$, and $10^{12}$
 cm, respectively. The base of the ejected shells correspond to a mass
 coordinate of $-3.05 \times 10^{-6}$. \label{fig2}}
\end{figure}

\clearpage

\begin{figure}
\plotone{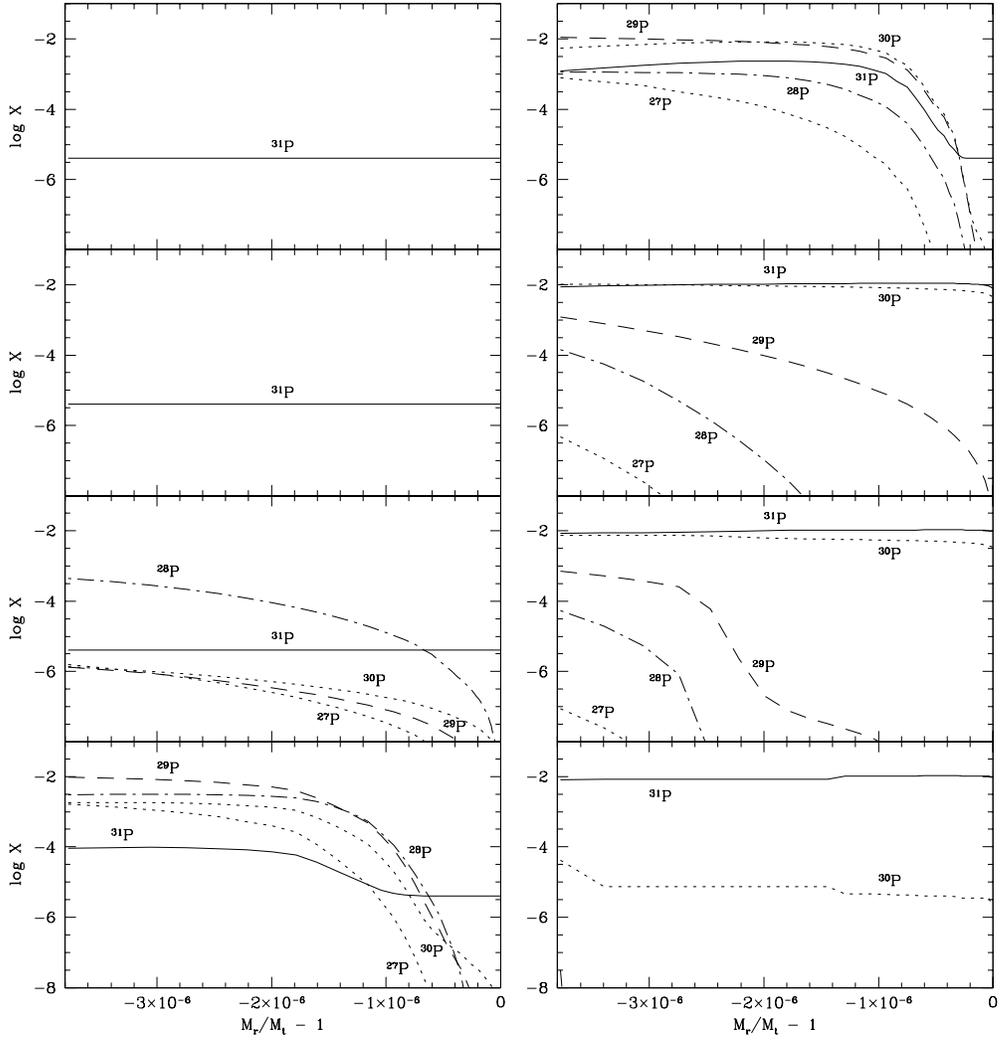}
\caption{Same as Fig. 2, for phosphorus isotopes. \label{fig3}}
\end{figure}

\clearpage

\begin{figure}
\plotone{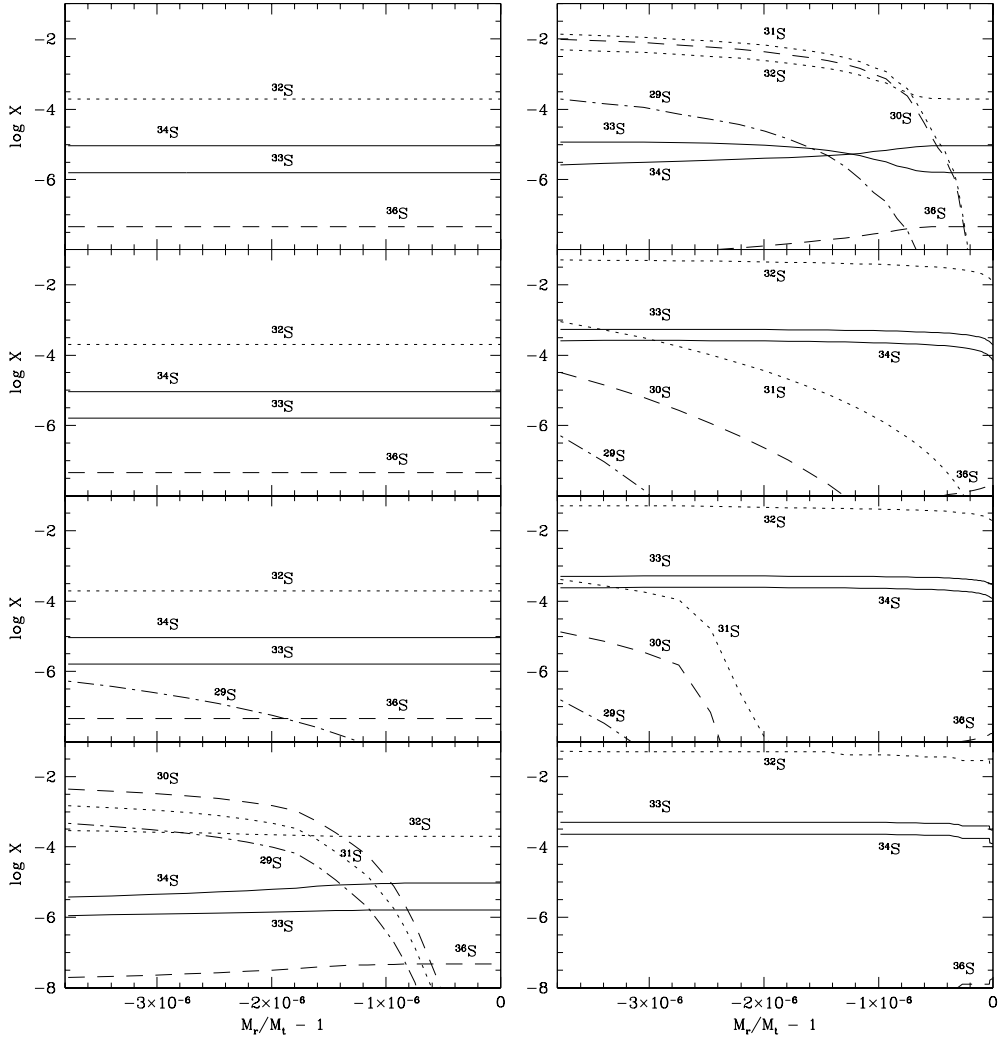}
\caption{Same as Fig. 2, for sulfur isotopes. \label{fig4}}
\end{figure}

\clearpage

\begin{figure}
\plotone{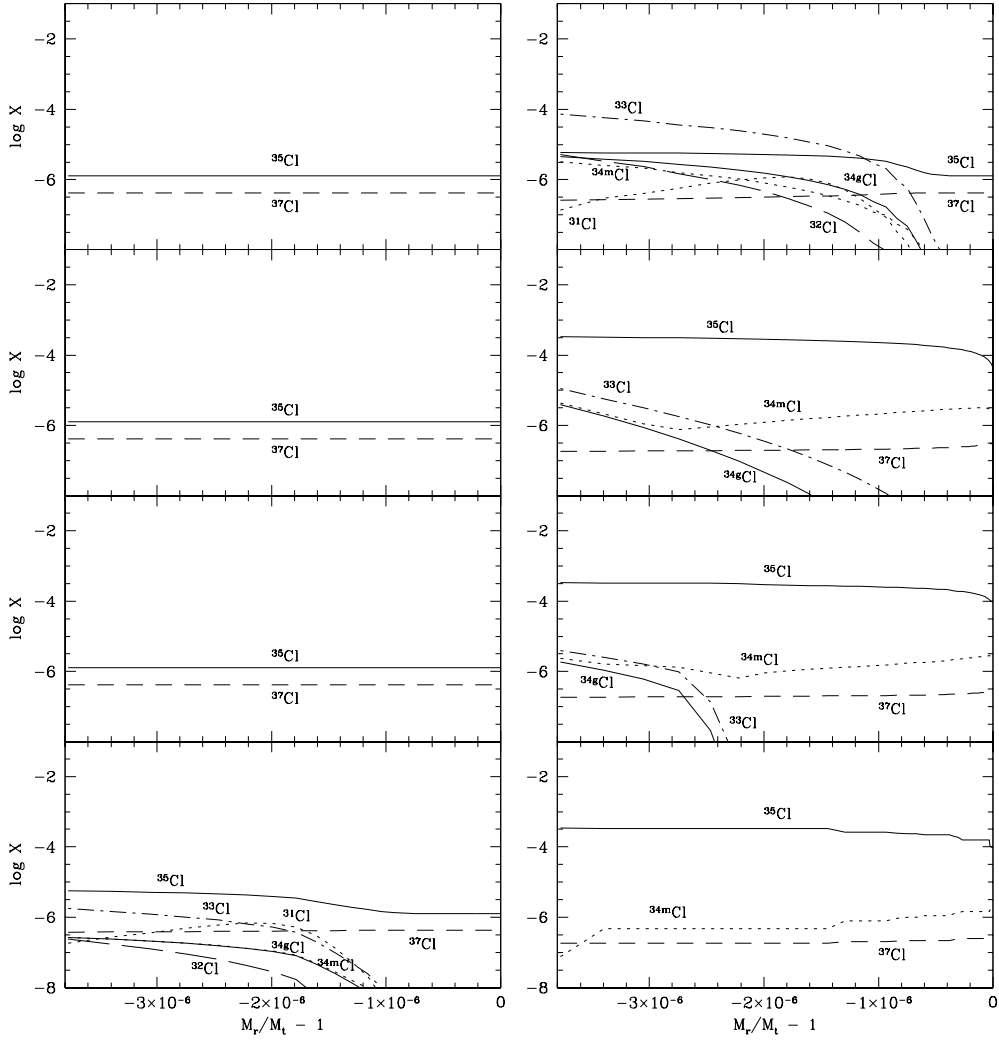}
\caption{Same as Fig. 2, for chlorine isotopes. \label{fig5}}
\end{figure}

\clearpage

\begin{figure}
\plotone{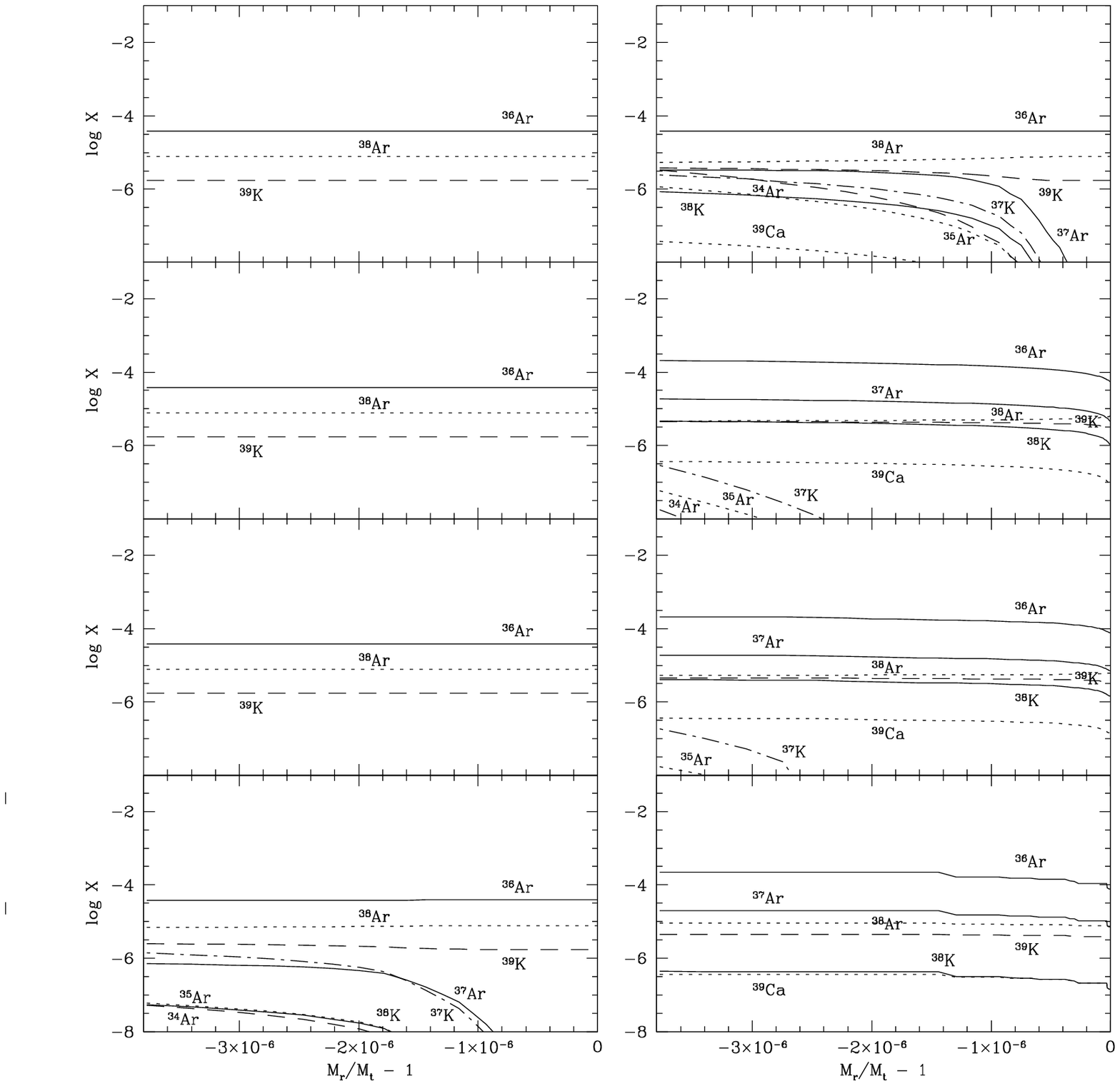}
\caption{Same as Fig. 2, for argon, potassium and 
       calcium isotopes. \label{fig6}}
\end{figure}

\clearpage

\begin{table}
\begin{center}
\caption{Initial composition (Si to Ca) in the accreted envelope, assuming
         a 50\% degree of mixing with the ONe white dwarf core (see Ritossa
         et al. 1996).} 

\begin{tabular}{cc} 
\tableline\tableline
Nuclei      & Mass fraction   \\
\hline
 $^{28}$Si  & 3.3E-4  \\
 $^{29}$Si  & 1.7E-5  \\
 $^{30}$Si  & 1.2E-5  \\
 $^{31}$P   & 4.1E-6  \\
 $^{32}$S   & 2.0E-4  \\
 $^{33}$S   & 1.6E-6  \\
 $^{34}$S   & 9.3E-6  \\
 $^{36}$S   & 4.7E-8  \\
 $^{35}$Cl  & 1.3E-6  \\
 $^{37}$Cl  & 4.3E-7  \\
 $^{36}$Ar  & 3.9E-5  \\
 $^{38}$Ar  & 7.7E-6  \\
 $^{39}$K   & 1.7E-6  \\
\tableline
\end{tabular}
\end{center}
\end{table} 

\clearpage

\begin{table}
\begin{center}
\caption{Mean composition of the ejecta (Si to Ca) from 1.35 \msun ONe white 
      dwarf models. Solar values are given for comparison.} 

\begin{tabular}{cccccc} 
\tableline\tableline
Nuclei      & Solar   &  Model   &  Model   & Model  & Model   \\
            &         &  ONe135B &p30low    & 135nom  & p30high \\
\hline
 $^{28}$Si  & 6.5E-4  &  4.2E-2  & 3.9E-2  & 4.4E-2 & 4.5E-2 \\
 $^{29}$Si  & 3.5E-5  &  2.1E-3  & 1.9E-3  & 1.9E-3 & 1.9E-3 \\
 $^{30}$Si  & 2.4E-5  &  1.4E-2  & 6.8E-2  & 1.4E-2 & 4.6E-4 \\
 $^{31}$P   & 8.2E-6  &  8.4E-3  & 1.1E-3  & 9.3E-3 & 1.2E-2 \\
 $^{32}$S   & 4.0E-4  &  2.3E-2  & 4.6E-3  & 4.5E-2 & 5.5E-2 \\
 $^{33}$S   & 3.3E-6  &  7.6E-3  & 4.6E-5  & 4.8E-4 & 5.9E-4 \\
 $^{34}$S   & 1.9E-5  &  9.3E-3  & 2.2E-5  & 2.2E-4 & 2.7E-4 \\
 $^{36}$S   & 6.4E-8  &  6.6E-9  & 5.6E-9  & 6.0E-9 & 6.1E-9 \\
 $^{35}$Cl  & 3.5E-6  &  4.5E-3  & 2.9E-5  & 2.9E-4 & 3.6E-4 \\
 $^{37}$Cl  & 1.2E-6  &  1.4E-4  & 9.0E-6  & 1.7E-5 & 2.0E-5 \\
 $^{36}$Ar  & 7.7E-5  &  2.2E-3  & 5.1E-5  & 1.9E-4 & 2.3E-4 \\
 $^{38}$Ar  & 1.5E-5  &  2.7E-5  & 8.1E-6  & 9.1E-6 & 9.5E-6 \\
 $^{39}$K   & 3.5E-6  &  5.2E-6  & 4.8E-6  & 4.7E-6 & 4.7E-6 \\
\tableline
\end{tabular}
\end{center}
\end{table} 

\end{document}